Nanometer-precision surface metrology of millimeter-size stepped objects using full-cascade-linked synthetic-wavelength digital holography using a line-by-line full-mode-extracted optical frequency comb

Eiji Hase,[1,4] Yu Tokizane,[1,4] Kazuki Sadahiro,[2] Takeo Minamikawa,[1] Isao Morohashi,[3] and Takeshi Yasui[1,*]

[1]Institute of Post-LED Photonics (pLED), Tokushima University, Tokushima, Tokushima 770-8506, Japan

[2]Graduate School of Sciences and Technology for Innovation, Tokushima University, Tokushima, Tokushima 770-8506, Japan

[3]National Institute of Information and Communications Technology, 4-2-1 Nukui-Kitamachi, Koganei, Tokyo 184-8795, Japan

[4]These authors contributed equally to this article.

*yasui.takeshi@tokushima-u.ac.jp

## Abstract

Digital holography (DH) is a powerful tool for surface profilometry of objects with sub-wavelength precision. In this article, we demonstrate full-cascade-linked synthetic-wavelength DH (FCL-SW-DH) for nanometer-precision surface metrology of millimeter-size stepped objects. 300 modes of optical frequency comb (OFC) with different wavelengths are sequentially extracted at a step of mode spacing from a 10GHz-spacing, 3.72THz-spanning electro-optic modulator OFC (EOM-OFC). The resulting 299 synthetic wavelengths and a single optical wavelength are used to generate a fine-step wide-range cascade link covering within a wavelength range of 1.54 μm to 29.7 mm. We determine the 0.1000mm-stepped surface with axial uncertainty of 6.1 nm within the maximum axial range of 14.85 mm.

# 1. Introduction

Non-contact, remote, surface profilometry is a powerful tool for quality control of various products in the industry. In particular, with recent advances in micromachining and microfabrication technologies, there is a growing demand for nanometer-precision surface metrology of millimeter-size objects. Among a variety of optical surface profilometry, digital holography (DH) [1-4] benefits from phase-image-based three-dimensional (3D) imaging with sub-wavelength axial precision, digital focusing, real-time measurement, and quantitative analysis by making full use of optical amplitude and phase images reconstructed from digital hologram. Its remaining drawback for the nanometer-precision surface metrology of millimeter-size objects is the maximum axial range (MAR) limited within a range of half the wavelength ($\lambda/2$) in a reflection configuration. This is because the optical phase repeats every $\lambda/2$ and suffers from phase wrapping ambiguity. Although a phase unwrapping analysis [5] has been often used for smooth surfaces with wavelength-order unevenness, it is difficult to apply it for stepped surfaces over multiple wavelengths. On the other hand, synthetic wavelength DH (SW-DH) [6-10], which uses two or three different wavelengths of light, increases the MAR and enables stepped surface metrology. However, the MAR is still limited within a range of a few tens μm; it cannot satisfy the requirement for the nanometer-precision surface metrology of millimeter-size objects.

Recently, multicascade-linked SW-DH (MCL-SW-DH) [11] has appeared to extend the MAR over millimeter. Use of a wavelength-tunable CW laser phase-locked to an optical frequency comb (OFC) enables to precisely synthesize a single optical wavelength and 5 synthetic wavelengths with different orders, and then seamlessly link them to achieve the MAR of 2.0 mm while securing the axial resolution of 34 nm. However, the need for sequential phase-locking operations at multiple optical wavelengths prevents the rapid imaging. To overcome this problem, line-by-line mode-extracted OFC was used for MCL-SW-DH in which 5 OFC modes with different wavelengths were sequentially extracted from a 10GHz-spacing electro-optic modulator OFC (EOM-OFC) instead of the tunable CW laser phase-locked to the OFC [12]. While the rapid switching of OFC mode could be repeated at 500 ms by using a combination of 2D spatial disperser and a spatial light modulator, cascaded links using a minimal number of synthetic wavelengths (= 4) and a single optical wavelength are sometimes vulnerable to the phase noise and cause the error to determine the number of phase wrapping because of coarse-step wide-range cascade links. To enhance the robustness to phase noise, the further increased number of cascade links is required.

In this article, we demonstrate full-cascade-linked SW-DH (FCL-SW-DH) by using ultra-multiple OFC modes available at a 10GHz-spacing, 3.72THz-spanning EOM-OFC. 299 synthetic wavelengths within a range of 99.4 µm to 29.7 mm are generated by sequentially extracting 300 OFC modes from the EOM-OFC at a step of mode spacing ($f_{rep}$). The resulting fine-step wide-range cascade links of synthetic wavelengths and optical wavelength largely enhances the robustness to phase noise while maintaining the large MAR and the low axial resolution.

## 2. Principle of operation

The synthetic wavelength ($\Lambda$) between two different optical wavelengths ($\lambda_1$, $\lambda_2$) is given by

$$\Lambda = \frac{\lambda_1 \lambda_2}{|\lambda_2 - \lambda_1|}. \tag{1}$$

When the EOM-OFC includes OFC modes within a range of mode number from $m$ to $m+\Delta m$ (wavelength = $\lambda_m$, $\lambda_{m+1}$, $\lambda_{m+2}$, •••, $\lambda_{m+i}$, •••, $\lambda_{m+\Delta m-1}$, $\lambda_{m+\Delta m}$) as shown by Fig. 1(a), $\Delta m$ different synthetic wavelengths can be generated by using different mode pairs of the EOM-OFC ($\lambda_m \& \lambda_{m+1}$, $\lambda_m \& \lambda_{m+2}$, •••, $\lambda_m \& \lambda_{m+i}$, •••, $\lambda_m \& \lambda_{m+\Delta m-1}$, $\lambda_m \& \lambda_{m+\Delta m}$).

$$\Lambda_i = \frac{\lambda_m \lambda_{m+i}}{|\lambda_{m+i} - \lambda_m|}, \tag{2}$$

where $i$ is the difference of mode number between two OFC modes ranging from 1 to $\Delta m$. In this way, it is possible to generate ultra-multiple synthetic wavelengths (= $\Lambda_1 > \Lambda_2 > ••• > \Lambda_i > ••• > \Lambda_{\Delta m-1} > \Lambda_{\Delta m}$) by sequentially extracting a single OFC mode from the EOM-OFC. These synthetic wavelengths covering a wide range of wavelength at a fine step as ultra-multiple cascades [see the green arrow in Fig. 1(b)] can be used for FCL-SW-DH together with an optical wavelength $\lambda$ [see the blue arrow in Fig. 1(b)].

We next consider the surface profilometry in a reflection configuration based on phase images with those ultra-multiple synthetic wavelengths (= $\Lambda_1$, $\Lambda_2$, •••, $\Lambda_i$, •••, $\Lambda_{\Delta m-1}$, $\Lambda_{\Delta m}$) and a single optical wavelength (= $\lambda$). The spatial distribution of height, $H(x, y)$, in a sample is given by

$$H(x,y) = \left[\frac{\phi_{\Lambda_1}(x,y)}{2\pi}\right]\frac{\Lambda_1}{2} = \left[N_{\Lambda_2}(x,y) + \frac{\phi_{\Lambda_2}(x,y)}{2\pi}\right]\frac{\Lambda_2}{2} = \cdots = \left[N_{\Lambda_{\Delta m-1}}(x,y) + \frac{\phi_{\Lambda_{\Delta m-1}}(x,y)}{2\pi}\right]\frac{\Lambda_{\Delta m-1}}{2}$$
$$= \left[N_{\Lambda_{\Delta m}}(x,y) + \frac{\phi_{\Lambda_{\Delta m}}(x,y)}{2\pi}\right]\frac{\Lambda_{\Delta m}}{2} = \left[N_\lambda(x,y) + \frac{\phi_\lambda(x,y)}{2\pi}\right]\frac{\lambda}{2},$$
(3)

where $N_{\Lambda i}(x, y)$ or $N_\lambda(x, y)$ is the spatial distribution of the phase wrapping number (integer) at $\Lambda_i$ or $\lambda$, and $\phi_{\Lambda i}(x, y)$ or $\phi_\lambda(x, y)$ is the spatial distribution of measured phase values at $\Lambda_i$ or $\lambda$. When the longest synthetic wavelength $\Lambda_1$ is set to indicate no phase wrapping, an unwrapped phase image $\phi_{\Lambda 1}(x, y)$ is obtained at $\Lambda_1$; $\phi_{\Lambda 1}(x, y)$ is used to calculate $H_{\Lambda 1}(x, y)$, where $H_{\Lambda 1}(x, y)$ is $H(x, y)$ determined by $\Lambda_1$. Then, $H_{\Lambda 1}(x, y)$ is used to determine $N_{\Lambda 2}(x, y)$. Subsequently, the determined $N_{\Lambda 2}(x, y)$ and the measured $\phi_{\Lambda 2}(x, y)$ are used to determine $H_{\Lambda 2}(x, y)$ more precisely than $H_{\Lambda 1}(x, y)$. Here, $N_{\Lambda i}(x, y)$ is given by

$$N_{\Lambda_i}(x,y) = INT\left[\frac{H_{\Lambda_{i-1}}(x,y)}{\Lambda_i/2} - \frac{\phi_{\Lambda_i}(x,y)}{2\pi}\right]. \quad (4)$$

By repeating a similar procedure from the longest synthetic wavelength ($\Lambda_1$) to the shortest synthetic wavelength ($\Lambda_{\Delta m}$) to the optical wavelength ($\lambda$), $N_\lambda(x, y)$ can be determined correctly. By using such a full-cascade link from $\Lambda_1$ to $\lambda$, both the maximum axial range of $\Lambda_1/2$ and the minimum axial resolution equal to phase noise (typically, $\lambda/10 \sim \lambda/100$) can be achieved at the same time. The resulting axial dynamic range becomes several orders of magnitude larger than the previous single SW-DH.

To clarify the difference between the present FCL-SW-DH and the previous MCL-SW-DH, cascade links of MCL-SW-DH are shown in Fig. 1(c) [12]. A minimal number of synthetic wavelengths (= 4) and a single optical wavelength with different orders of wavelength (=$\Lambda_1$, $\Lambda_2$, $\Lambda_3$, $\Lambda_4$, and $\lambda$) generated coarse-step wide-range cascades of wavelength. In this case, high accuracy of phase measurement is required to determine the phase wrapping number at each synthetic wavelength and optical wavelength [$N_{\Lambda_i}(x, y)$ and $N_\lambda(x, y)$] without errors. On the other hand, in the fine-step wide-range cascades of wavelength in FCL-SW-DH [see Fig. 1(b)], the requirement for phase measurement accuracy is largely relaxed. In other words, the FCL-SW-DH is more robust to the phase noise than the MCL-SW-DH discussed later.

### 3. Materials and methods

3.1 Experimental setup

Figure 2(a) shows a schematic drawing of experimental setup. We constructed the experimental setup of FCL-SW-DH composed of an EOM-OFC (center wavelength = 1550 nm, frequency spacing = 10 GHz corresponding to wavelength spacing of 0.08 nm, spectral bandwidth = 3.72 THz corresponding to 30 nm, total output power = 500 mW), a single-mode fiber (SMF), a polarization controller (PC), a tunable ultra-narrowband optical bandpass filter (BPF; Alnair labs, CVF-300CL, center wavelength = 1525~1610 nm, optical passband = 3.7~370 GHz corresponding to 20~3000 pm, insertion loss = 5.5 dB), a variable optical attenuator (VOA; Thorlabs, Inc., V1550A, wavelength = 1250~1650 nm, maximum attenuation = 25dB), and an off-axis Michelson interferometer. The detail of the EOM-OFC is given in the previous paper [12-14]. 300 OFC modes with moderate optical power were sequentially extracted from the EOM-OFC by BPF. Since optical power of each extracted OFC mode is different among them due to complicated spectral shape, it was adjusted to maintain at a constant level (= 1µW) by VOA.

In the optical setup of the off-axis Michelson interferometer, an object beam and a reference beam were separated by a beam splitter cube (BS, reflection = 50%, transmittance = 50%). The object beam was reflected at a sample whereas the reference beam was reflected at a reference mirror (surface roughness = 63.2 nm). After their reflected beams were combined again by the same BS, they were incident on a Peltier-cooled infrared CCD camera (Allied Vision Technol. GmbH, Goldeye P-008, 320 × 256 pixels, exposure time = 10 ms, digital output resolution = 14 bit) at an off-axis angle of 1º to generate interference patterns with a fringe spacing of 100 µm, namely, the digital hologram. We repeated the acquisition of 300 digital holograms with different optical wavelengths. An angular spectrum method (ASM) [15,16] was used to reconstruct an optical phase image from the digital hologram. We generated 299 synthetic-wavelength phase images by using different pairs of 300 optical-wavelength phase images with different wavelengths. Finally, we performed full-cascade linking of 299 synthetic-wavelength phase images and a single optical-wavelength phase image [See Fig. 1(b)].

3.2 Sample

To investigate the effectiveness of the FCL-SW-DH for the nanometer-precision surface metrology of millimeter-size stepped objects, we prepared a metal sample having known stepped surface. Figures 2(b) and 2(c) show an optical photograph and a schematic drawing of the sample, respectively. The sample was composed of four steel gauge blocks with different thicknesses (Mitutoyo, 516-458; thickness = 8.0000±0.00014 mm, 2.0000±0.00014 mm, 1.100±0.00014 mm, and 1.0000±0.00014 mm), which were tightly attached on an optical flat (Edmund Optics Inc., 43-416-000, surface flatness = 31.7 nm) by wringing. The resulting four surfaces has a height of 8.0000±0.00014 mm [surface (A)], 2.0000±0.00014 mm [surface (B)], 1.1100±0.00014 mm [surface (C)], and 1.0000±0.00014 mm [surface (D)] from the optical flat surface. The resulting step difference are 7.0000±0.00020 mm for step (A-D), 6.9000±0.00020 mm for step (A-C), 6.0000±0.00020 mm for step (A-B), 0.9000±0.00020 mm for step (B-C), and 0.1000±0.00020 mm for step (C-D).

## 4. Results and discussion

4.1 Basic performance of line-by-line full-mode-extracted EOM-OFC

Since the minimum optical passband of BPF (= 3.7 GHz) is sufficient to extract a single OFC mode from the 10-GHz spacing EOM-OFC, we evaluated the extracted OFC mode. We measured optical spectra of EOM-OFC before BPF and the extracted single OFC mode after BPF by an optical spectrum analyzer (Yokogawa Test & Measurement Corp., Tokyo, Japan, AQ-6315A, wavelength range = 350–1750 nm, wavelength resolution = 0. 05 nm). Figures 3(a) shows an optical spectrum before BPF, corresponding to that of the EOM-OFC. The EOM-OFC had a center wavelength of 1550 nm and covered a wavelength range of 1541.92~1566.22 nm, corresponding to a center optical frequency of 193.414 THz and an optical frequency range of 191.412~194.428 THz. Fine modulation of the spectral envelope reflects a series of discrete, equally 10GHz-spaced OFC modes of EOM-OFC. Sequential single-mode extraction from these OFC modes enables FCL-SW-DH. Figure 3(b) shows an optical spectrum of the extracted single OFC mode at 1548.673 nm. Neighboring OFC modes at both sides completely disappeared. We selected 30 continuous OFC modes within a wavelength range of 1547.00~1549.45 nm from extracted 300 OFC modes for comparison. Figure 3(c) compares optical spectra of those 30 OFC modes at a step of $f_{rep}$. A series of $f_{rep}$-interval-tunable single OFC modes with a constant power level was obtained for FCL-SW-DH.

We also calculated available synthetic wavelengths generated by line-by-line full mode extraction of the 10-GHz-spacing, 3.72-THz-spanning EOM-OFC. Figure 4 shows the synthetic wavelength with respect to the difference of mode number ($\Delta m$) between two OFC modes when the synthetic wavelength is sequentially generated by increasing $\Delta m$ with an increment of *1*. Use of 300 OFC modes enables us to generate 299 synthetic wavelengths within a range of 99.4 µm to 29.7 mm.

4.2 Axial uncertainty and its spatial unevenness with respect to the number of cascade links

To evaluate the effectiveness of the increased number of cascade links in

FCL-SW-DH, we determined the step difference from cascade-linked phase images with respect to the number of cascade links. We here selected the three steps with different orders of step difference: step (A-C) with a step difference of 6.9000±0.00020 mm, step (B-C) with a step difference of 0.9000±0.00020 mm, and step (C-D) with a step difference of 0.1000±0.00020 mm. 300 digital holograms with different optical wavelengths were acquired without image accumulation and then were used for cascade liking among SW-DHs. We obtained $H(x, y)$ for each surface from its cascade-linked phase images. Then, we calculated the step difference from the mean and standard deviation of $H(x, y)$ for two surfaces that make up the step. Red, blue, and green plots in Fig. 5(a) show the measured step difference with respect to the number of cascade links for steps (A-C), (B-C), and (C-D), respectively. For comparison, specification values of those three steps are indicated as red, blue, and green lines in Fig. 5(a). We confirmed that the measured step difference converged to the specification value depending on the increased number of cascade links.

We next calculated the axial uncertainty of step (C-D) with a step difference of 0.1000 mm [see green plot and line in Fig. 5(a)] when the axial uncertainty was defined as the difference of step difference between the measured value and the specification value. Figure 5(b) shows a relation between the number of cascade links and the axial uncertainty for step (C-D). As the number of cascade links increases, the axial uncertainty decreases from micrometer order to nanometer order. The final link from the shortest synthetic wavelength to the optical wavelength achieves the axial uncertainty of 6.1 nm.

We also investigated the spatial unevenness when it is defined as a standard deviation of axial uncertainty. Red, blue, and green plots in Fig. 5(c) show a relation between the number of cascade links and the spatial unevenness for steps (A-C), (B-C), and (C-D), respectively. The spatial unevenness decreased depending on the number of cascade links for all of steps (A-C), (B-C), and (C-D) similar to the axial uncertainty in Fig. 5(b). The final link to the optical wavelength achieves the spatial unevenness of 376 nm for step (A-C), 218 nm for step (B-C), 56 nm for step (C-D), respectively. In this way, the increased number of cascade links enhances the axial uncertainty and the spatial unevenness, suppressing the error in determining the number of phase wrapping.

4.3 3D shape measurement of sub-millimeter and millimeter stepped surfaces

Finally, we perform the 3D shape measurement of the stepped surface sample [see Figs 2(b) and 2(c)] by the FCL-SW-DH with 299 synthetic wavelengths and a single optical wavelength. Figure 6(a) shows the 3D visualization of the sample. 4 stepped surfaces with sub-millimeter and millimeter step differences were clearly visualized. Each surface indicates little errors in the determination of phase wrapping number in each cascade link. This result confirms the good axial accuracy and spatial unevenness shown in Fig. 5.

To highlight the advantage of FCL-SW-DH over the previous MCL-SW-DH [12], we perform the 3D shape measurement of the same sample by MCL-SW-DH

using 4 synthetic wavelengths (= 29720.66 µm, 14860.33 µm, 1238.360 µm, and 99.40000 µm) and a single optical wavelength (= 1.541924 µm). Figure 6(b) shows the 3D visualization of the stepped surface. Although 4 stepped surfaces with sub-millimeter and millimeter step differences were visualized, errors in determining the phase wrapping number appears as discrete unevenness of height. The coarse-step wide-range cascade links with the limited number of synthetic wavelengths and a single optical wavelength makes the MCL-SW-DH vulnerable to phase noise, resulting in degraded axial accuracy compared with the FCL-SW-DH. In contrast, the increased number of cascade links in FCL-SW-DH increases the robustness to phase noise and hence enhances the axial uncertainty due to the fine-step wide-range cascade links with ultra-multiple synthetic wavelengths and a single optical wavelength.

## 5. Conclusions

We demonstrate FCL-SW-DH for nanometer-precision surface metrology of millimeter-size objects. Sequential line-by-line full mode extraction of the 10GHz-spacing, 3.72THz-spanning EOM-OFC enables the 299-step cascade links within the wavelength region of 1.54 µm to 29.7 mm, extending to the MAR up to 14.85 mm. The effectiveness of the FCL-SW-DH was highlighted in the profilometry of a stepped metal surface. The fine-step wide-range cascade links suppressed errors in determination of the phase wrapping number and enhances the axial uncertainty down to 6.1 nm and the spatial height unevenness down to 56 nm. The comparison of the stepped surface profilometry between the FCL-SW-DH and the previous MCL-SW-DH clearly indicates the superiority of the former in nanometer-precision surface metrology of millimeter-size objects. The FCL-SW-DH will be a powerful tool for surface topography of various objects in industry.

**Acknowledgements.** The authors thank Drs. Yasuaki Hori and Akiko Hirai at the National Institute of Advanced Industrial Science and Technology (AIST) for their help in the measurement of the gauge-block sample.

**Funding.** Cabinet Office, Government of Japan (Subsidy for Reg. Univ. and Reg. Ind. Creation).

**Disclosures.** The authors declare no conflicts of interest.

**Data availability.** Data underlying the results presented in this paper are not publicly available at this time but may be obtained from the authors upon reasonable request.

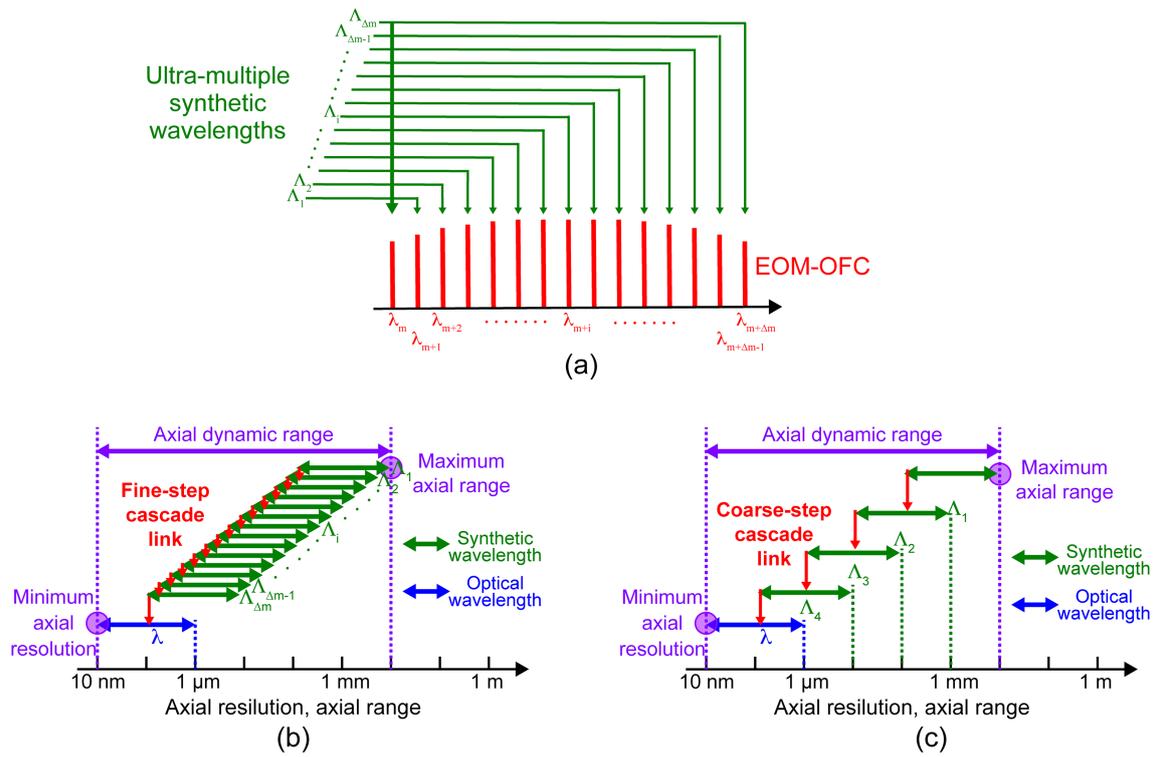

Fig. 1. Principle of operation. (a) Generation of ultra-multiple synthetic wavelengths by line-by-line full-mode-extracted EOM-OFC. (b) Fine-step wide-range cascade links of wavelength in FCL-SW-DH. (c) Coarse-step wide-range cascade links of wavelength in MCL-SW-DH.

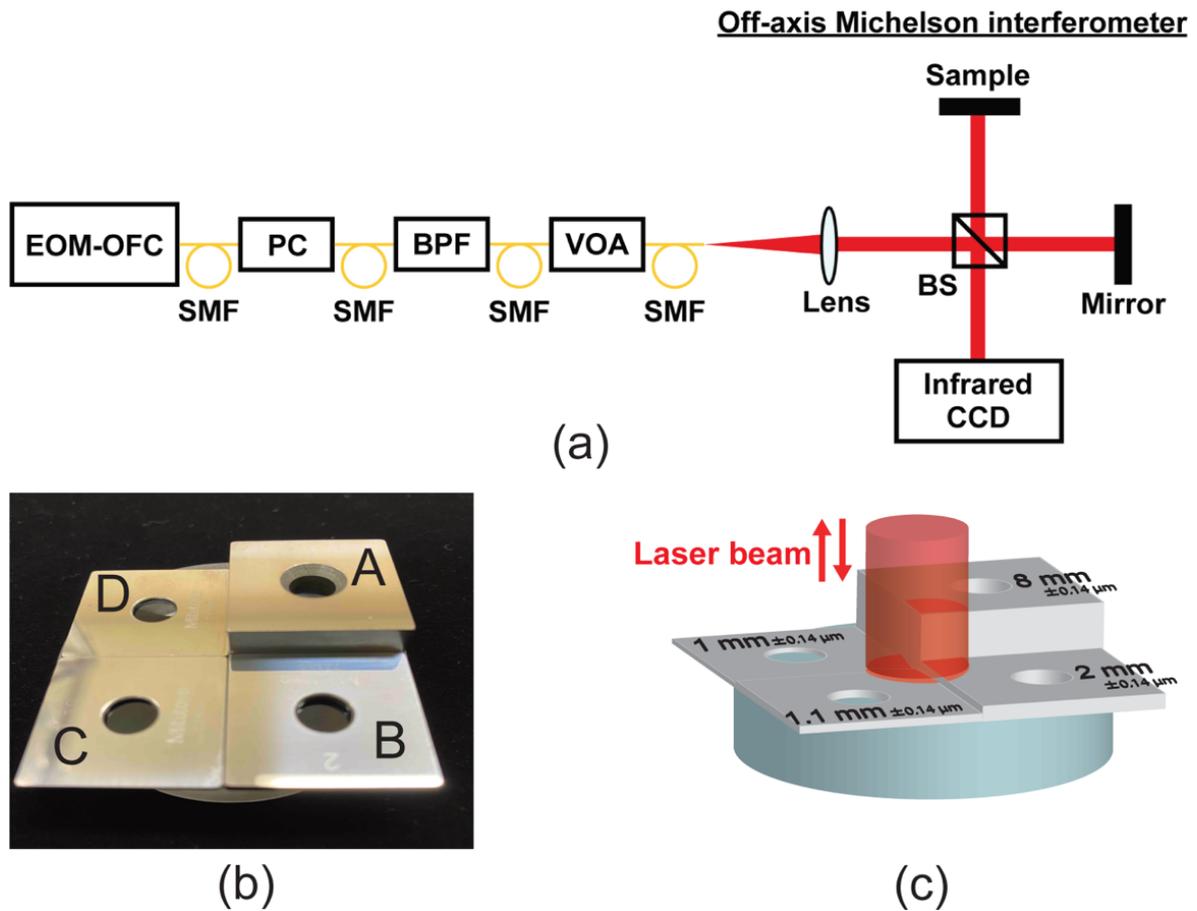

Fig. 2. (a) Experimental setup. EOM-OFC, electro-optic modulator optical frequency comb; SMF, single-mode fiber; PC, polarization controller; BPF, tunable ultra-narrowband optical bandpass filter; VOA, variable optical attenuator; BS, beam splitter cube. (b) Optical photograph and (c) schematic drawing of a stepped surface sample.

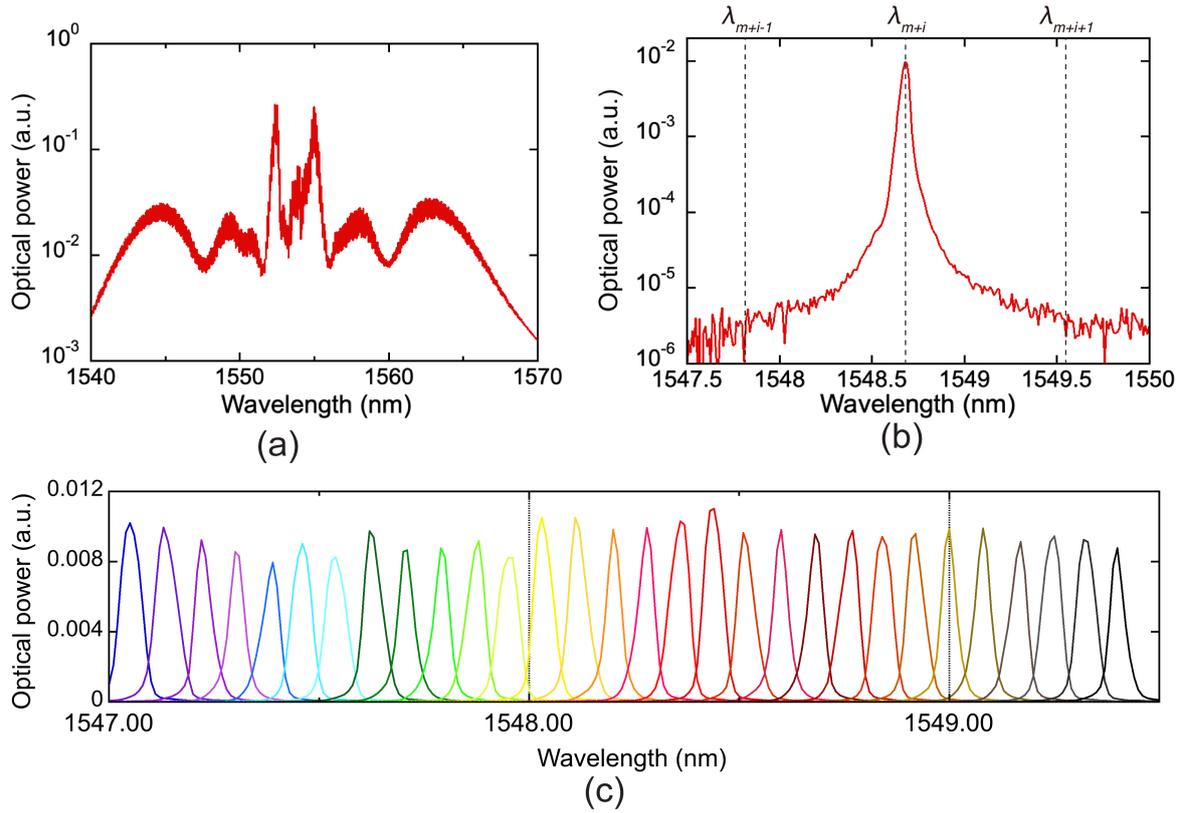

Fig. 3. (a) Optical spectrum of EOM-OFC. Optical spectra of (b) a single OFC mode and (c) 30 continuous OFC modes generated by a line-by-line full-mode-extracted EOM-OFC.

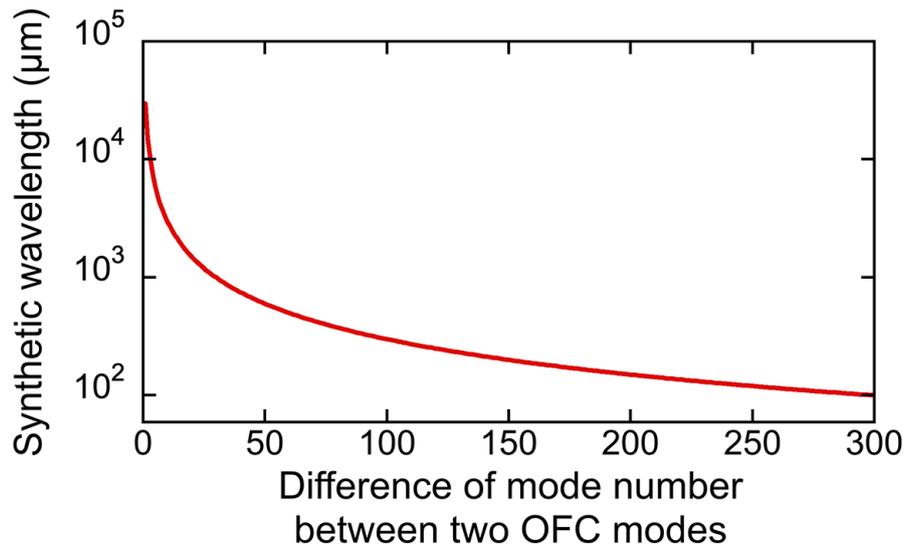

Fig. 4. Relation between the difference of mode number (Δ*m*) between two OFC modes and synthetic wavelength.

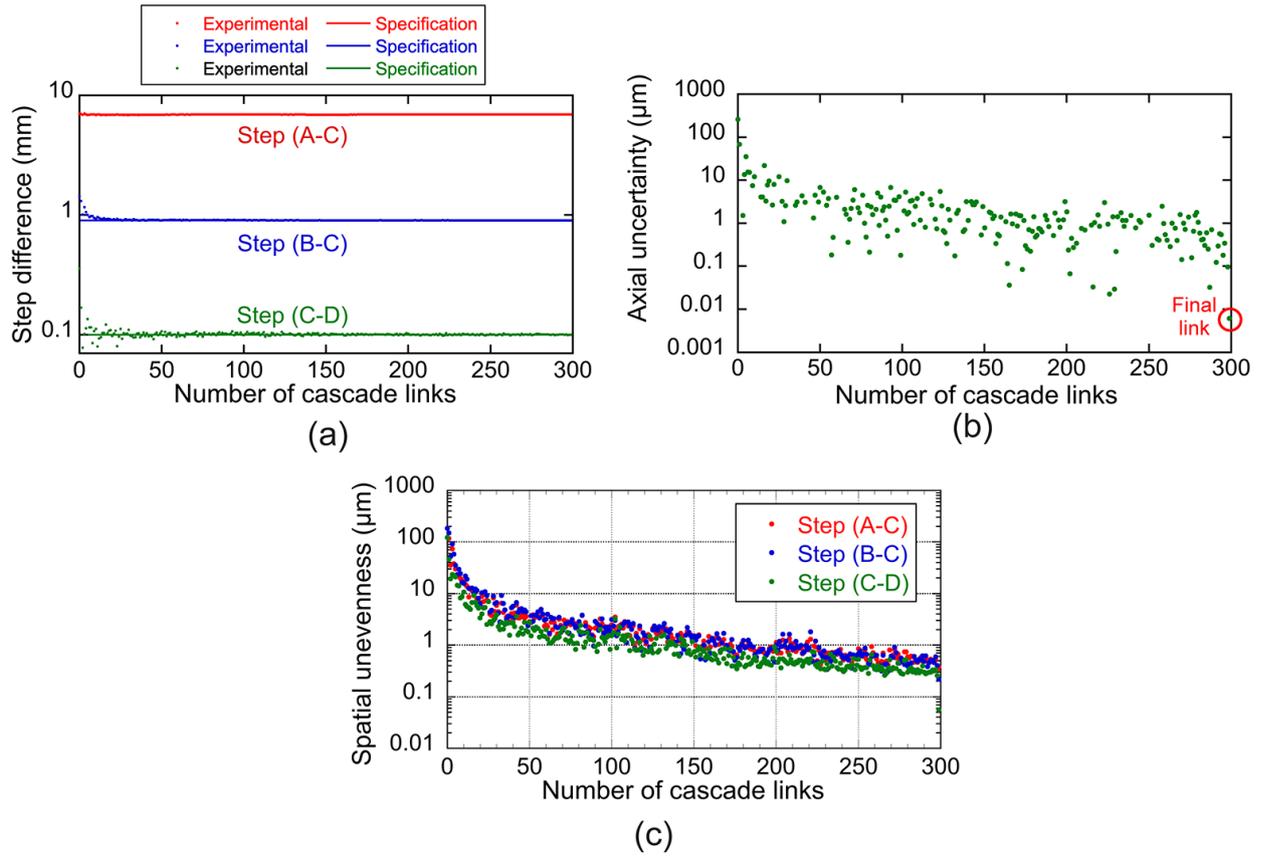

Fig. 5. (a) Measured step difference with respect to the number of cascade links for steps (A-C), (B-C), and (C-D), respectively. (b) Relation between the number of cascade links and the axial uncertainty for step (C-D). (c) Relation between the number of cascade links and the spatial unevenness for steps (A-C), (B-C), and (C-D), respectively.

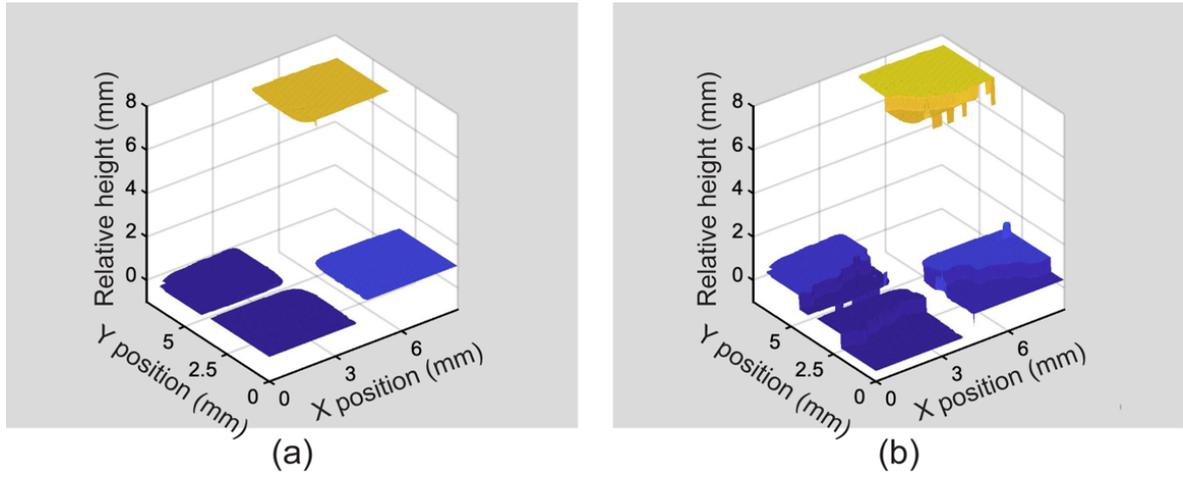

Fig. 6. 3D shape measurement of sub-millimeter and millimeter stepped surfaces measured by (a) the present FCL-SW-DH and (b) the previous MCL-SW-DH.